\documentclass[aps,twocolumn,floats,prd,nofootinbib,superscriptaddress,10pt]{revtex4-2}
\usepackage[british]{babel}
\usepackage[dvipsnames]{xcolor}
\usepackage{booktabs}
\usepackage{latexsym}
\usepackage{amssymb}
\usepackage{amsmath}
\usepackage{graphicx}
\usepackage{tabularx}
\usepackage{cancel}
\usepackage{hyperref}

\hypersetup{
    colorlinks=true,
    linkcolor=red,
    citecolor=blue,
    filecolor=black,
    urlcolor=blue
}
\usepackage[capitalise]{cleveref}
\usepackage[utf8]{inputenc}
\usepackage[normalem]{ulem}
\def\beq#1\eeq{\begin{align}#1\end{align}}

\newcommand{\dd}{\text{d}}
\newcommand{\ee}{\text{e}}

\definecolor{darkgreen}{RGB}{0,170,0}

\begin{document}

\newcommand{\UniCa}{\affiliation{Dipartimento di Fisica, Universit\`a di Cagliari, Cittadella Universitaria, 09042 Monserrato, Italy}}
\newcommand{\INFNCa}{\affiliation{INFN, Sezione di Cagliari, Cittadella Universitaria, 09042 Monserrato, Italy}}
\newcommand{\INFNROMA}{\affiliation{INFN, Sezione di Roma, Piazzale Aldo Moro 2, 00185, Roma, Italy}}
\newcommand{\Sapienza}{\affiliation{Dipartimento di Fisica, Sapienza Università 
	di Roma, Piazzale Aldo Moro 5, 00185, Roma, Italy}}
\newcommand{\inafoac}{\affiliation{INAF - Osservatorio Astronomico di Cagliari, Via della Scienza 5, 09047 Selargius (CA), Italy}}
\newcommand{\gssi}{\affiliation{
Gran Sasso Science Institute (GSSI), Viale F. Crispi 7, L'Aquila (AQ), I-67100, Italy}}
\newcommand{\infngssi}{\affiliation{INFN - Laboratori Nazionali del Gran Sasso (LNGS), L'Aquila (AQ), I-67100, Italy}}

\author{Mariano~Cadoni}
\email{mariano.cadoni@ca.infn.it}
\UniCa\INFNCa

\author{Andrea~Cozzumbo}
\email{andrea.cozzumbo@gssi.it}
\gssi\infngssi

\author{Riccardo~Murgia}
\email{riccardo.murgia89@unica.it}
\UniCa\INFNCa\inafoac

\author{Matteo~Pili}
\email{matteo.pili@inaf.it}
\inafoac\UniCa\INFNCa

\author{Mirko~Pitzalis}
\email{mirko.pitzalis@ca.infn.it}
\UniCa\INFNCa

\author{Andrea~P.~Sanna}
\email{asanna@roma1.infn.it}
\Sapienza\INFNROMA

\author{Matteo~Schulz}
\email{matteo.schulz@gssi.it}
\gssi\infngssi

\author{Ilaria~Usai}
\email{ilaria.usai@ca.infn.it}
\UniCa\INFNCa

\title{Late-time cosmic acceleration {from matter inhomogeneities}: constraints from DESI DR2 and SNe Ia}

\begin{abstract}
We present the first observational constraints on the Misner-Sharp (MS) mass of the universe, within a newly proposed exact General Relativity solution providing a non-perturbative
description of nonlinear cosmic structure formation. At early times, when statistical homogeneity and isotropy are satisfied, our model reduces to usual FLRW cosmology, accommodating Cosmic Microwave Background (CMB) data. At late times, the departure from homogeneity and isotropy on scales $\lesssim 200$~Mpc -- codified in the scale-dependent redshift distribution of the MS mass
-- naturally induces cosmic acceleration, through an extensive term generated by the MS mass asymptotic limit on large scales.
We derive the MS mass from Baryon Acoustic Oscillation data from the Dark Energy Spectroscopic Instrument (DESI) and Type Ia Supernovae, obtaining a value of the cosmological constant in agreement with $\Lambda$CDM results. We outline broader implications and future observational prospects, highlighting how multi-probe strategies -- particularly CMB lensing and the late Integrated Sachs-Wolfe effect -- can ultimately test this solution to the dark energy puzzle.
\end{abstract}

\maketitle
%\tableofcontents

{\bf \em Introduction$-$} The standard $\Lambda$ Cold Dark Matter ($\Lambda\text{CDM}$) {cosmological} paradigm faces profound theoretical tensions~\cite{Carroll:1991mt,Carroll:2000fy,Bull:2015stt,CosmoVerseNetwork:2025alb}. While it successfully fits background expansion data, it does so at the cost of the vacuum energy problem~\cite{Peebles:2002gy,Martin:2012bt,Calcagni:2022ssd},~\textit{i.e.},~the $\mathcal{O}(10^{60})$ gap between observed cosmic acceleration and quantum field theory predictions, and the coincidence problem~\cite{Zlatev:1998tr,Velten:2014nra},~\textit{i.e.},~the mystery of why late-time acceleration occurs precisely when nonlinear structures become dominant. In fact, the need to introduce a cosmological constant $\Lambda$ was clear well before the detection of cosmic acceleration~\cite{Perlmutter:1997zf,Riess:1998cb,Perlmutter:1998np}, to correctly reproduce the observed large-scale structure (LSS) properties with results from CDM cosmological $N$-Body simulations~\cite{Blumenthal:1984bp,Efstathiou:1985zz,Davis:1985rj,White:1987zz,Frenk:1988zz,Maddox:1990hb,Efstathiou:1990xe,Ostriker:1995su,KraussTurner1995}.
To effectively describe the observed LSS clustering properties, $N$-Body simulations assuming $\Lambda$CDM rely on an \textit{a priori} imposed background geometry, where dark energy (DE) and LSS growth are entirely decoupled by construction~\cite{Peebles:1980yev,Mukhanov:1990me}.
In the absence of an exact analytical solution unifying nonlinear structure formation and cosmic acceleration, these conceptual flaws have been tolerated as the price for a remarkably good phenomenological fit~\cite{Hogg:2004vw,Yadav:2005vv,Maartens:2011yx,2013MNRAS.429.2910C,Perivolaropoulos:2014lua,Pandey:2015xea,Bengaly:2016amk,Bengaly:2017slg,Bengaly:2018ykb,Secrest:2020has,Krishnan:2021jmh,Dias:2023qrl,Lopez:2024kih,Sawala:2025ajg,Labini:2025dnc}. However, the unprecedented precision reached by LSS surveys finally enables testing the foundational $\Lambda\text{CDM}$ assumptions, such as the late-time statistical homogeneity and isotropy~\cite{Blomqvist:2008ud,shafieloo_clarkson_2010,Colin:2010ds,Scrimgeour:2012wt,Alonso:2014xca,Camarena:2021mjr,Aluri:2022hzs,Dam:2022wwh,Secrest:2022uvx,Tang:2023kzs,Bengaly:2024ree,Hu:2024qnx,Seifert:2024bqr,Courtois:2025xcs,Heinesen:2026wwy,Koksbang:2026jxb,Koksbang:2026wvh}, and the interpretation of DE as a cosmological constant~$\Lambda$~\cite{Wu:1998ad,Linder:2005in,Copeland:2006wr,Brax:2017idh,CANTATA:2021asi,DESI2024VI,DESIDR2II,DESI:2024aqx,DESI:2025fii,DESI:2025ffa,Afroz:2025iwo,Scherer:2025esj,Mukherjee:2025ytj,Gallavotti:2026lja,Gallavotti:2026och}. Building upon a companion paper~\cite{Cadoni:2026etd} that establishes its exact analytical foundation within General Relativity (GR), in this \textit{Letter} we introduce the Anisotropic Fluid Inhomogeneous Cosmology (AFIC) framework, which addresses both aforementioned $\Lambda$CDM cracks at once.
The core novelty lies in abandoning the standard perturbative assumption of small metric fluctuations, enabling a fully relativistic treatment of non-linear cosmic structure growth.
Within AFIC, early-universe statistical homogeneity and isotropy naturally approaches the usual Friedmann-Lema\^itre-Robertson-Walker (FLRW) predictions, recovering Big Bang Nucleosynthesis (BBN)~\cite{Cyburt:2015mya} and Cosmic Microwave Background (CMB)~\cite{Planck:2018vyg} constraints. In contrast, at late times, due to nonlinear structure growth, the spacetime geometry departs from the homogeneous and isotropic FLRW limit. This departure is described by an exact solution to Einstein's equations. 
Subject to the appropriate boundary conditions, the geometry is determined by a single input function, $M(r,z)$,~\emph{i.e.},~the scale-dependent redshift distribution of the Misner--Sharp (MS) mass\footnote{The MS mass~\cite{Misner:1964je} is a purely
geometric quantity, fully covariant. In a flat universe, it measures the total mass enclosed within a sphere with areal radius $R=r/(1+z)$ at a given redshift $z$.}. Its deviation from a purely homogeneous FLRW contribution on intermediate and small scales ($\lesssim 200$~Mpc) encodes the evolution of nonlinear matter inhomogeneities~\cite{Cadoni:2026etd}. 
However, on sufficiently large scales the AFIC metric and field equations reduce to flat FLRW cosmology. Our predictions can therefore be tested with standard observational analyses of cosmic expansion probes. The crucial difference compared to $\Lambda\text{CDM}$ is that the large-scale asymptotic behavior of the MS mass $M(r \to \infty, z)$ intrinsically connects nonlinear LSS growth and emergence of cosmic acceleration at late times, offering a solution to the coincidence problem.
AFIC provides a concrete, analytical realization of a historic cosmological conjecture, entirely formulated within GR: that the nonlinear growth of LSS can dynamically generate late-time cosmic acceleration. Progress along this path was hindered by 
the analytical complexity of spatial averaging schemes, the presence of backreaction terms, and ambiguities related to the nonlinearity of Einstein's field equations~\cite{Ellis:1984bqf,Ellis:1987zz,Buchert2000,Buchert2001,Kolb:2005da, Buchert:2006ya,Wiltshire2007b,Green:2010qy,Green:2013yua,Clarkson:2011zq,Clarkson:2011uk,Sussman:2011na,Wiltshire:2011vy,BuchertRasanen2012,Racz:2016rss,Buchert:2018yhd,Buchert:2019mvq,Clifton:2019cep,Brunswic:2020bjx,Breton:2020puw,Lapi:2023plb,Giani:2024nnv,Gomes:2024iyq,Raffai:2025ldr,Galoppo:2025rkd,Fleury:2025ykl,Galoppo:2026rin}, leaving the physical viability of inhomogeneous cosmological models heavily debated~\cite{Baumann:2010tm, Bolejko:2015gmk,Adamek:2014gva,Adamek:2017mzb,Macpherson:2018btl,Macpherson:2021gbh,Macpherson:2022eve,Camarena:2022iae,Oestreicher:2024mtw,Adamek:2024hme,Camarena:2025upt,Adamek2026Comments}. AFIC resolves these limitations~\cite{Cadoni:2026etd} converting a longstanding hypothesis into a fully falsifiable framework. If high-precision multi-probe observations will demonstrate that {the only input function} $M(r, z)$ matches the observed expansion history, late-time cosmic acceleration will be established as a structural GR effect. Should it not occur, the AFIC model will help pointing to where and how alternative extended frameworks must step in.
This \textit{Letter} focuses on providing a first observational determination of the only physical input of  the AFIC solution, the MS mass. To this end, we make use of Baryon Acoustic Oscillation (BAO) data from DESI DR2~\cite{DESIDR2II} and the catalog of Type Ia Supernovae (SNe Ia) from Pantheon+~\cite{Scolnic:2021amr,Brout:2022vxf}. From the MS mass we derive the value of the effective DE-like component, in agreement with the $\Lambda$CDM value of the cosmological constant.  We also highlight how synergies among CMB lensing and the late Integrated Sachs Wolfe (ISW) effect, could allow us to determine~$M(r, z)$ directly from observations. This would make AFIC fully predictive, enabling the cosmology community to establish whether the late-time cosmic acceleration can be explained without postulating exotic forms of DE and/or extending GR.

{\bf \em Theoretical framework$-$} 
The full theoretical and mathematical derivation of AFIC is presented in detail in Ref.~\cite{Cadoni:2026etd}. Here we summarize the core formulas necessary for observational testing. We {effectively} describe matter {in the universe} with an anisotropic fluid. We adopt the metric ansatz~\cite{Cadoni:2024jxy,Cadoni:2024rri,Cadoni:2023lum,Cadoni:2023lqe,Cadoni:2022chn,Cadoni:2021zsl,Cadoni:2020izk,Cadoni:2018dnd} (unless explicitly stated, we use natural units, namely $c = 1$): 
\begin{equation}
\label{metric}
\dd s^2=a^2(t)\left[{-\ee^{\alpha(t,r)}}\dd t^2+{\ee^{\beta(t,r)}}\dd r^2+r^2\dd\Omega^2\right]\, ,
\end{equation}
with $\alpha$ and $\beta$ functions of the conformal time $t$ and the comoving radial coordinate $r$  and $a(t)$ the scale factor.
Note that we are assuming spherical symmetry and radial density profiles, which may be a good approximation mostly to describe voids. Rather than as an ultimate description of the inhomogeneous universe,~\cref{metric} should be regarded as a useful tool to explore the interplay between matter and geometry in regimes which cannot be described by the oversimplified, fully symmetric FLRW metric.
Einstein's field equations can be written as follows~\cite{Cadoni:2026ejk,Cadoni:2026etd} (dots and primes denote derivatives with respect to $t$ and $r$, respectively)
\begin{subequations}
\label{eq:IntegrteFields0}
\begin{align}
\label{IntegrteFields.a}
&\ee^{-\beta} = 
1 - \frac{2  G M}{R}
+ \frac{8\pi  G}{3} R^2   {\rho}_{\text{\tiny FLRW}}\ee^{-\alpha}\,; \\
\label{IntegrteFields.b}
&\frac{\dot a}{a}\alpha' +\frac{\dot \beta}{r}=0\, ;  \\
\label{IntegrteFields.c}
&\rho =  \frac{1}{4\pi a^3 r^2} 
\partial_r \!M\, ; \\
\label{IntegrteFields.d}
&p_{\parallel} = -  \frac{1}{4\pi \dot a a^2 \, r^3} 
\partial_t M\, ; \\
\label{IntegrteFields.e}
&p_{\perp} = 
p_{\parallel} 
+ \frac{r}{2} \partial_r p_{\parallel} 
+ \frac{1}{4} r \alpha' \left(\rho + p_{\parallel}\right)\, .
\end{align}
\end{subequations}
$R \equiv ra$ is the areal radius; $\rho(r, t) \equiv \tilde\rho_{\text{\tiny FLRW}}(t)+\rho_{\rm I}(r, t)$ is the total energy density of the universe~\cite{Cadoni:2026etd}; {$\rho_{\text{\tiny FLRW}}(t)=\tilde \rho_{\text{\tiny FLRW}}(t)+ \rho_{\rm I}(r \to \infty, t)$} is the asymptotic limit of $\rho(r, t)$; $p_\perp$ and $p_{\parallel}\equiv \tilde{p}_{\parallel, \text{\tiny FLRW}}(t)+ p_{\parallel, \rm I}(r, t)$ are the tangential and radial pressures, respectively. $M(r,t)$ denotes the MS mass, which is the only free input function. In the large-scale (or early-time) limit, \cref{metric} becomes the standard FLRW metric\footnote{Note that at late times, on intermediate and small scales, where the metric effectively deviates from FLRW, \cref{metric} might violate the Copernican principle (see the discussion in~Ref.~\cite{Cadoni:2026etd}). However, as current constraints on such violation~\cite{Zhang:2010fa,Bull:2011wi} are derived within different spacetime geometries or with perturbative approaches assuming $\Lambda\text{CDM}$, they do not apply to AFIC. We leave their reformulation to future work.}, and \cref{eq:IntegrteFields0} reduces to the standard Friedmann equations for an isotropic and barotropic fluid \cite{Cadoni:2026etd}. 
Once $M(r,t)$ is specified, the system of~\cref{IntegrteFields.a,IntegrteFields.b,IntegrteFields.c,IntegrteFields.d,IntegrteFields.e} admits a unique solution~\cite{Cadoni:2026ejk}. Such a solution must be supplemented by appropriate boundary conditions~\cite{Cadoni:2026etd} to enforce homogeneity and isotropy on large scales,~\textit{i.e.},~in the $r \to \infty$ limit. At late times, when $\rho_{\rm I}(r, t)$ significantly departs from zero due to nonlinear structure formation, the dynamics of the homogeneous background cannot be decoupled from that of perturbations, as described by~\cref{{eq:IntegrteFields0}}.
By relying on the full, exact gravitational dynamics, our approach avoids the difficulties inherent in alternative frameworks based on averaging and backreaction~\cite{Ellis:1984bqf,Ellis:1987zz,Buchert2000,Buchert2001,Buchert:2006ya,Wiltshire2007b,Green:2010qy,Green:2013yua,Clarkson:2011zq,Clarkson:2011uk,Sussman:2011na,Wiltshire:2011vy,BuchertRasanen2012,Racz:2016rss,Buchert:2018yhd,Buchert:2019mvq,Clifton:2019cep,Brunswic:2020bjx,Breton:2020puw,Lapi:2023plb,Giani:2024nnv,Gomes:2024iyq,Galoppo:2025rkd,Fleury:2025ykl,Galoppo:2026rin}.
The splitting of the density induces an analogous separation of the MS mass\footnote{Hereafter we trade the scale factor $a(t)$ for the redshift $z$.} as $M(r,\, z) = M_\text{\tiny FLRW}\left(R,\tilde \rho_{\text{\tiny{FLRW}}}\right) + M_\text{I}(r,\, z)$, where ${M_\text{\tiny FLRW}} \equiv \left(4\pi/3\right) R^3 \tilde \rho_\text{\tiny FLRW}$ accounts for the FLRW homogeneous contribution, while $M_\text{I}(r,\, z)$ encodes the contribution arising from inhomogeneities in the matter distribution. Since in the early universe $M_\text{I}(r,\, z)\ll M_\text{\tiny FLRW}$, it can be treated with standard linear perturbation theory on an FLRW background. 
Its spatial averaging is expected to give the required tiny fraction of energy density at recombination to accommodate BBN and CMB data~\cite{Cyburt:2015mya,Planck:2018vyg}. In contrast, to recover the large-scale FLRW limit at late times, we must necessarily have 
\begin{equation}
\label{eq:mlimit}
    M(r, \, z)
\sim \frac{4\pi}{3} R^3 \rho_{\text{\tiny FLRW}}  \quad \text{for} \quad r \to \infty\, ,
\end{equation}
with $\rho_{\text{\tiny FLRW}} \equiv \tilde{\rho}_{\text{\tiny FLRW}}+\lambda \, g(z)$. We assume $\tilde \rho_{\text{\tiny FLRW}}$ to be that of non-relativistic (baryonic and dark) matter,~\textit{i.e.}, $\tilde \rho_{\text{\tiny FLRW}}(z) \propto (1+z)^3$. We stress, however, the need to go beyond the standard collisionless CDM paradigm, as CDM alone cannot be the source of the large anisotropic stress required to explain late-time cosmology. The second contribution to $\rho_{\text{\tiny FLRW}}$ comes from the $r\to\infty $ limit of $M_\text{I}(r,\, z)$, whose details are contained in the function $g(z)$, while $\lambda$ is a constant with dimensions of $\text{length}^{-4}$. As such, $\lambda \, g(z) \equiv \rho_{\rm I}(r \to \infty, z)$ represents the contribution to the large-scale dynamics coming from matter inhomogeneities. While within $\Lambda$CDM this term is attributed to a cosmological constant, within AFIC it plays the role of an effective emergent DE component generated by matter itself. Although being a quite unnatural hypothesis, we cannot, however, exclude the usual interpretation of  the term $\lambda g(z)$: namely, the assumption that it describes an exotic DE component already present in the early universe, yet manifesting itself only at late times (within AFIC as an extensive term in the MS mass asymptotic limit). Despite these different possible interpretations, our model might accommodate the results reported by DESI without extending GR or introducing exotic fluids. The mechanism is quite general, as the asymptotic behavior of the MS mass does not depend on its specific functional form~\cite{Cadoni:2026etd}. Not entering the metric solution, $\lambda \, g(z)$ does not contribute to the local gravitational potential \cite{Cadoni:2026etd}: it emerges at late times as a large-scale effect induced by the nonlinear growth of matter inhomogeneities on smaller scales ($\lesssim 200$~Mpc). To obtain the same effect, standard homogeneous cosmology assumes that DE does not cluster and affects only large scales at late times.

{\bf \em Implications and observational prospects$-$}
Rather than relying on effective equation-of-state parametrizations~\cite{Chevallier:2000qy,Linder:2002et} or alternative gravity theories~\cite{Carroll:2003wy,Carroll:2004de,Amendola:2006we,Amendola:2006kh,DeFelice:2010aj,Clifton:2011jh,Joyce:2014kba,Ratra:1987rm,Caldwell:1997ii,Carroll:1998zi,Tsujikawa:2013fta,Horndeski:1974wa,Deffayet:2011gz}, AFIC model's single input function, $M(r,z)$, links late-time cosmic expansion to the non-linear LSS geometry. This is a simple theoretical alternative, grounded within GR, which we now test by determining the asymptotic value of $M(r,z)$ using state-of-the-art cosmic expansion data.
To do so, we build upon Refs.~\cite{Hellaby:2006cj,Lu:2007gr} where it was demonstrated that measuring the maximum of the areal radius $ R_{\text{max}}$ on our past null cone is equivalent to measuring the mass within a sphere of radius $R_{\text{max}}$~\cite{Cadoni:2026etd}. The latter is reached at the apparent horizon~\cite{Hellaby:2006cj} so that, in a flat universe, the total MS mass inside a sphere with areal radius $R_{\text{max}}$ is (hereafter we reinstate the units of $c$)
\begin{equation}\label{hellaby}
M(R_{\text{max}},z_{\text{max}}) = \frac{R_{\text{max}}c^2}{2G}\,
\end{equation}
$R_{\text{max}}$ is in fact the only point on the past null cone allowing for a direct, model-independent estimate of the total cosmic mass~\cite{Hellaby:2006cj}. \Cref{hellaby} is quite general, as it requires spherical symmetry, not statistical homogeneity.
We now determine $R_{\text{max}}$ from angular diameter distance $D_A(z)$ and luminosity distance $D_L(z)$ data provided by BAO and SNe Ia, respectively\footnote{As AFIC is entirely formulated within GR, the distance duality relation~\cite{Etherington:1933asu} ensures a one-to-one correspondence between $D_A(z)$ and $D_L(z)$.}, exploiting the fact that $R_{\text{max}} \equiv D_A(z_\text{max})$. Determining $M(R_{\text{max}},z_{\text{max}})$ via \cref{hellaby} allows us to estimate, within AFIC, the effective DE density:
\begin{equation}\label{eq:rho}
\rho_{\rm I}(R_{\text{max}},z_{\text{max}}) = \frac{3c^2}{8\pi G}\frac{1}{R_{\text{max}}^2} - 
{\tilde \rho}_{\text{\tiny FLRW}}(z_{\text{max}})\, .
\end{equation}
Although completely model-independent, \cref{hellaby} does not provide a \textit{direct} determination of the MS mass.  Hence, in this work we set constraints on the total energy density 
$\rho(R_{\text{max}},z_{\text{max}})$, not on its individual contributions. {While confronting our results against the $\Lambda$CDM prediction~$\rho_{\text{\tiny FLRW}}(z_{\text{max}})$} represents a first, necessary, experimental check, it does not allow us to break the degeneracy between the two terms in the right-hand side of~\cref{eq:mlimit}.
In general, any alternative cosmological model seeking to replace $\Lambda\text{CDM}$ must provide a good fit to at least five key probes: $(i)$ CMB anisotropies and BBN;
$(ii)$ background expansion measurements from BAO and SNe Ia at $z \lesssim 2$;
$(iii)$ growth rate of structure on linear and quasi-linear scales;
$(iv)$ weak lensing data across $0 \lesssim z \lesssim 2$; $(v)$ amplitude of the late ISW effect ($A_{\text{ISW}}$) and CMB lensing.
AFIC satisfies requirement $(i)$ by construction, as in the absence of nonlinear inhomogeneities anisotropies in the stress-energy tensor are switched off, recovering standard linear perturbations on a FLRW background. This \textit{Letter} focuses on requirement $(ii)$. Following the strategy presented in Ref.~\cite{Hellaby:2006cj} and summarized in~\cref{eq:rho}, we extract ${R_{\text{max}}}$ from DESI BAO DR2~\cite{DESIDR2II} and Pantheon+ SNe Ia data~\cite{Scolnic:2021amr,Brout:2022vxf}, making no assumptions on the underlying cosmological model.
Perturbative analyses, including points $(iii)$ to $(v)$ above alongside background-growth consistency tests~\cite{Ishak:2013vha,Sen:2026zuc,Gao:2026uwo}, can be performed by implementing AFIC's full set of equations~\cite{Cadoni:2026etd,Cadoni:2020izk} within publicly available Einstein-Boltzmann solvers~\cite{Blas:2011rf,Zumalacarregui:2016pph,Bellini:2019syt,Cataneo:2024kfa} and/or relativistic $N$-Body codes~\cite{Adamek:2016zes,Lepori:2020ifz,Lepori:2021lck,Adamek:2026gev2}.

\begin{figure}[bthp]
\centering
\includegraphics[width=0.5\textwidth]{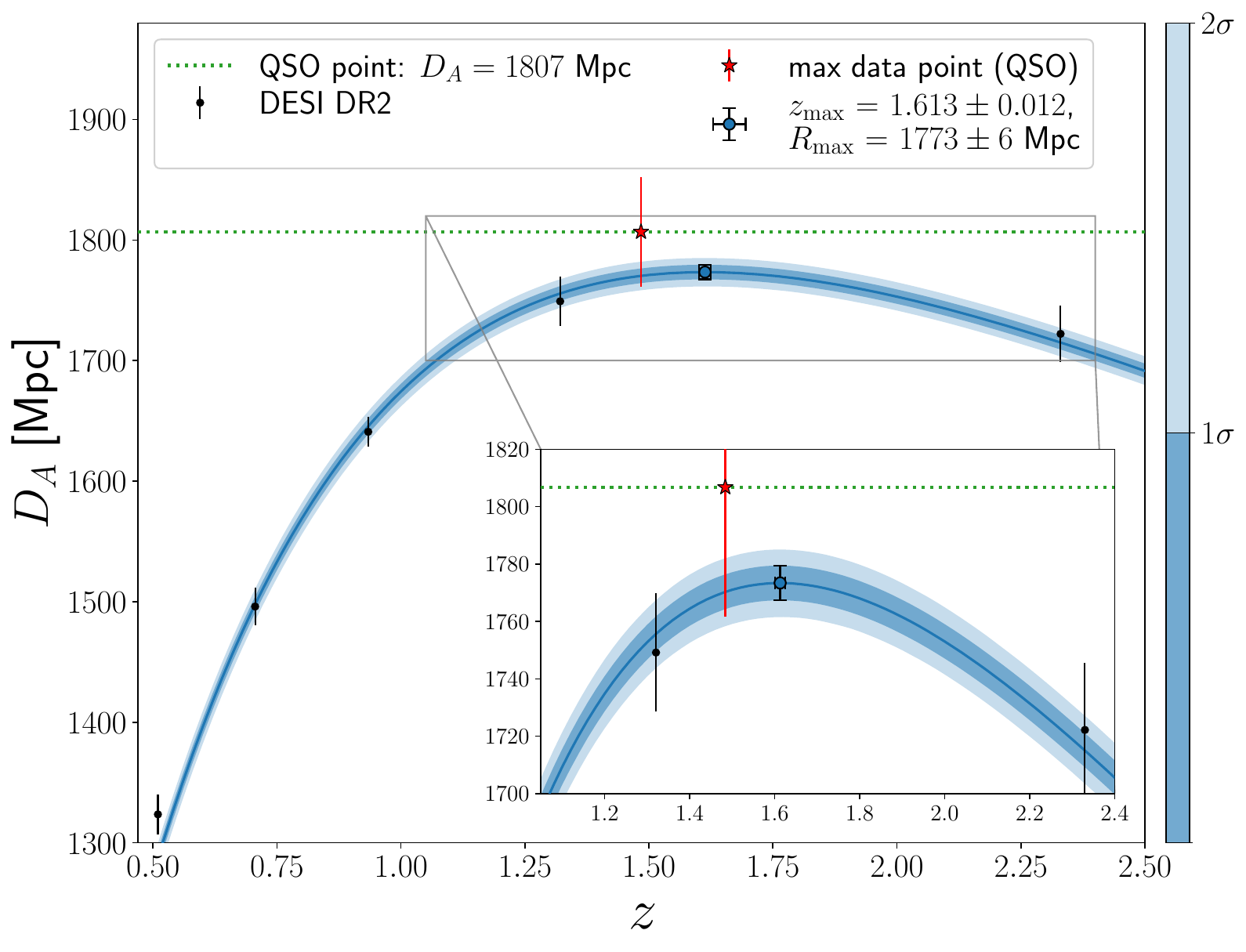}
\caption{\textit{Model-independent constraints on $R_{\max}$ from BAO}. Data points and error bars are from DESI DR2, assuming $r_d=147.09$~Mpc. The QSO point at $z=1.484$ is shown as a red star, with the corresponding value of $D_A(z)$ highlighted by the horizontal green dotted line. The blue solid curve is the median of our GP reconstruction; the blue circle identifies the resulting estimate of $\{R_{\max}$, $z_{\max
}\}$, with its 68\% uncertainty. The shaded bands correspond to 68\% and 95\% C.L.}
\label{fig:DA_DESI}
\end{figure}

\begin{figure}[bthp]
\centering
\includegraphics[width=0.5\textwidth]{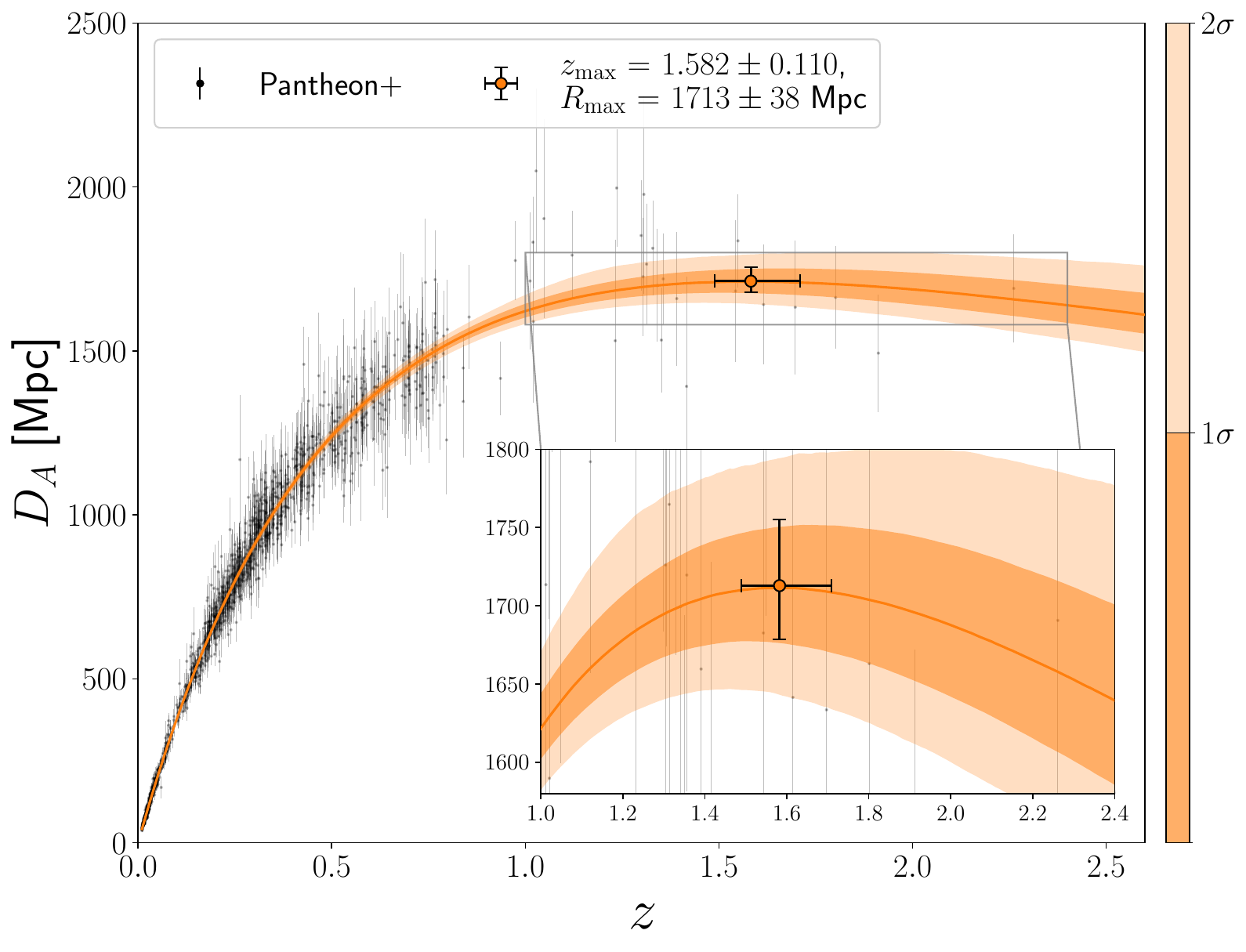}
\caption{\textit{Model-independent constraints on $R_{\max}$ from SNe Ia}.  Data points and error bars are from Pantheon+, assuming $h=0.7$. The red solid curve is the median of our GP reconstruction; the orange circle identifies the resulting estimate of $\{R_{\max}$,$z_{\max
}\}$, with its 68\% uncertainty. The shaded bands correspond to 68\% and 95\% C.L.}
\label{fig:DA_SN}
\end{figure}

{\bf \em Constraints from BAO and SNe Ia$-$} 
As BAO measure distances in units of $r_d$, namely the sound horizon at the baryon drag epoch\footnote{Changing $r_d$ would change MS mass estimate and $\Lambda$CDM prediction accordingly, with no impact on our conclusions.}, to express them in Mpc we assume $r_d=147.09\pm0.26$~Mpc~\cite{Planck:2018vyg}.
A direct, model-independent measurement of $D_A(z_{\max})$ would provide a fully accurate estimate of $R_{\max}$, and in turn, of the total energy density of the universe today,~\textit{i.e.},~$\rho(R_{\max},z_{\max})$. 
The most straightforward approach to obtain a rough estimate is to consider the quasar (QSO) data point from DESI DR2, at $z=1.484$ (red star in Figure~\ref{fig:DA_DESI}), being the closest data point to the standard FLRW expectation, that is~$ z_{\max} \approx 1.6$. Taking the QSO point yields $R_{\max} \simeq 1807$~Mpc. {Alternatively, one can
locate} $D_A(z_{\max})$ in BAO data with a Gaussian Process (GP) method~(see~\ref{SM:1}). 
Results are reported in Figure~\ref{fig:DA_DESI}.
We find $z_{\max}\simeq1.613$ and $R_{\max}\simeq1773$~Mpc, $M(R_{\max}, z_{\max}) \simeq 1.853\,\cdot\,10^{22}\,M_\odot$. The resulting estimate of the total energy density of the universe is $\rho(R_{\max},z_{\max}) = (53.68\pm0.37) \,\cdot\, 10^{-27}\,{\rm kg\,m^{-3}}$.
We now apply the GP method to Pantheon+ data~\cite{Scolnic:2021amr,Brout:2022vxf}. We fit distance moduli $\mu(z)$ of 1590 SNe Ia, where $D_A=10^{(\mu-25)/5}/(1+z)^2$~Mpc. The normalization\footnote{A different choice would change MS mass estimate and $\Lambda$CDM prediction accordingly, with no impact on our conclusions.} is set by $h=0.7$. Results are shown in Figure~\ref{fig:DA_SN}. We find $z_{\max}\simeq1.58$, $R_{\max}\simeq1713$~Mpc, $M(R_{\max}, z_{\max}) \simeq 1.79\,\cdot\,10^{22}\,M_\odot$. The total energy density is $\rho(R_{\max},z_{\max})=\left(57.5\pm2.6\right) \,\cdot\, 10^{-27}\,{\rm kg\,m^{-3}}$.~All our constraints are compatible with the flat $\Lambda$CDM prediction with matter abundance $\Omega_m \approx 0.3$ and dimensionless Hubble parameter $h \approx 0.7$, namely $\rho_{\rm{\scriptscriptstyle FLRW}}(z_{\max}) \approx 55 \,\cdot\, 10^{-27}\,{\rm kg\,m^{-3}}$~(see~\ref{SM:2}).

{\bf \em Discussion$-$}
The AFIC framework provides a minimal GR solution beyond the assumption of statistical homogeneity and isotropy underlying usual FLRW cosmology. Unlike standard cosmology, the gravitational dynamics is sourced by an anisotropic fluid. The framework is specified by a single, fully covariant input function: the MS mass $M(r,z)$, which encodes the scale-dependent redshift distribution of cosmic matter, and must ultimately be determined observationally. At early times, when statistical homogeneity and isotropy hold across all scales, AFIC approaches standard linear perturbation theory around an FLRW background. At late times, as nonlinear structure formation generates significant inhomogeneities and pressure anisotropies on small and intermediate scales, the FLRW solution is recovered only asymptotically, in the large-scale limit $r\to\infty$. Within AFIC, the emergence of cosmic acceleration is fully determined by the asymptotic behavior of the MS mass of the total matter content of the universe. This provides a dynamical mechanism to address the coincidence problem without introducing a cosmological constant or additional \emph{ad hoc} extra fields.
In this $Letter$, we have derived model-independent constraints on the MS mass of the universe on Gpc scales using two independent probes of the background expansion history: BAO from DESI DR2 and SNe Ia from Pantheon+. Both yield consistent estimates of the effective DE density. However, as illustrated above, background expansion probes alone cannot shed light on the true nature of the term $\lambda g(z)$. Breaking the degeneracy in \cref{eq:mlimit} and distinguishing emergent acceleration from exotic DE will require a direct, cosmology-independent reconstruction of the redshift evolution of the MS mass on scales $\gtrsim 200\text{ Mpc}$. A promising avenue is the cross-correlation of CMB temperature maps with galaxy catalogs, to isolate the amplitude of the late ISW effect ($A_{\mathrm{ISW}}$)~\cite{Crittenden:1995ak,Planck:2015qru}. Complementarily, high-resolution CMB lensing convergence maps from upcoming facilities like the Simons Observatory~\cite{Ade:2019owm} and CMB-S4~\cite{Abazajian:2019eic} will trace the integrated matter distribution. Combined with BAO and redshift-space distortion (RSD) data, these probes could enable us to reconstruct the MS mass across all relevant scales. A failure of the proposed relation would help identify which aspects of the underlying gravitational dynamics require extension beyond GR. Conversely, confirming a link between nonlinear LSS growth and cosmic expansion would establish late-time acceleration as an emergent feature of inhomogeneous GR solutions. 
If so, since standard collisionless CDM cannot generate the correct anisotropy level to fit observations, the AFIC model would yield a paradigmatic shift, converting the need of exotic DE to a criterion to select the most viable non-CDM scenarios.

{\bf \em Acknowledgements}$-$  We warmly thank Julian Adamek, Rodrigo Calderón, Matteo Cinus and Nils Sch$\ddot{\rm o}$neberg, for many insightful comments and suggestions which definitely improved this manuscript. We thank Eric V. Linder for useful discussions, as well as the organizers and participants in the \href{https://indico.kgwg.org/event/158/program}{Korea–Italy Workshop on Multi-Messenger Astrophysics and Cosmology 2026}, particularly Stefano Ascenzi, Ulyana Dupletsa, Benjamin L'Huillier, Gor Oganesyan, Arman Shafieloo.
APS acknowledges financial support by the MUR FIS2 Advanced Grant ET-NOW (CUP:~B53C25001080001). MC, RM, MP, MPit, APS, and IU acknowledge support by the INFN TEONGRAV specific initiative. We acknowledge support from the \href{https://www.sgws-community.eu}{Sardinian GW Science (SGWS) Community}: a joint program between the University of Cagliari and the Gran Sasso Science Institute (GSSI) promoted and coordinated by the cultural association \href{https://www.associazioneideas.eu}{IDeAS}.\\

\bibliography{refs}

\onecolumngrid
%\clearpage

\clearpage

\section*{Supplemental Material}
%----------------------------------------------------------------------------
\renewcommand{\thesubsection}{{SM.\arabic{subsection}}}
\setcounter{section}{0}

\subsection{Gaussian process reconstruction of $D_A(z)$ and $D_L(z)$}
\label{SM:1}

To locate the maximum of $D_A(z)$ without assuming a cosmological model, we reconstruct $D_A(z)$ directly from distance data with Gaussian processes (GP). Below we summarize the method, the choice of the GP mean, how $z_{\max}$ and $R_{\max}$ are extracted and the robustness tests we performed.

A GP is a probability distribution over functions. A function $f(z)$ drawn from a GP with mean $m(z)$ and kernel $k(z,z')$, $f\sim\mathcal{GP}(m,k)$, takes values at any set of redshifts $z_i$ that are jointly Gaussian, with means $m(z_i)$ and covariance $k(z_i,z_j)$. We mainly use the squared-exponential kernel
\begin{equation}
k(z,z')=\sigma_f^2\,\exp\!\left[-\frac{(z-z')^2}{2\ell_f^2}\right],
\end{equation}
whose two hyperparameters have a direct meaning: $\sigma_f$ is the typical amplitude of the deviations of $f$ from the mean, and $\ell_f$ is the redshift interval over which these deviations are correlated, i.e.\ how smooth the function is.
Conditioning the GP on the data, with their full covariance, gives the distribution of functions compatible with them, from which we draw realizations. The derivative of a GP is itself a GP, with covariance given by the derivatives of $k$, so each realization provides $f$ and $f'$ consistently~\cite{Hwang:2022hla}.

The choice of the mean matters~\cite{Hwang:2022hla}. A zero mean is not robust: away from the data the reconstruction relaxes towards zero, which produces spurious oscillations where the data are sparse. A mean fixed to a given model instead pulls the reconstruction towards that model and underestimates its errors. We therefore take the most agnostic mean, namely a flat background with dark-energy equation of state $w(a)=w_0+w_a(1-a)$~\cite{Chevallier:2000qy,Linder:2002et},
\begin{equation}
H^2(z)=H_0^2\left[\Omega_{\rm m}(1+z)^3+(1-\Omega_{\rm m})(1+z)^{3(1+w_0+w_a)}\,e^{-3w_a z/(1+z)}\right],
\end{equation}
where $\Omega_{\rm m}$ is the present matter density parameter and $H_0=100\,h$~km\,s$^{-1}$\,Mpc$^{-1}$, with $h=0.7$ fixed. The parameters $\Omega_{\rm m}$, $w_0$, $w_a$ are sampled together with $\sigma_f$ and $\ell_f$ and marginalized over.

The mean is thus not imposed: it is marginalized over many expansion histories, $\Lambda$CDM ($w_0=-1$, $w_a=0$) among them. All priors are flat: $\Omega_{\rm m}\in[0.05,0.8]$, $w_0\in[-3,1]$, $w_a\in[-3,2]$ and $\sigma_f\in[0,5]$, in units of $D_M/r_d$ for DESI and in magnitudes for Pantheon+. For $\ell_f$ the upper bound is $10$ for DESI and $5$ for Pantheon+, and the lower bound is set by the Nyquist--Shannon sampling theorem to twice the spacing $\Delta z$ of the redshift grid on which the GP is evaluated, the shortest scale that grid can resolve: $0.04$ for DESI ($\Delta z=0.02$) and $0.02$ for Pantheon+ ($\Delta z=0.01$). The posteriors of $\ell_f$ lie well above these bounds.

The function reconstructed by the GP is not $D_A$ itself, but the quantity closest to the data: $f=D_M/r_d$ for DESI and the distance modulus $\mu$ for Pantheon+. For each posterior sample we draw one realization of this function, conditioned on the data, and convert it to $D_A$: $D_A=f\,r_d/(1+z)$ for DESI and $D_A=10^{(\mu-25)/5}/(1+z)^2$~Mpc for Pantheon+. In each realization, $z_{\max}$ is identified as the redshift where $\dd D_A/\dd z$ changes sign from positive to negative, and $R_{\max}=D_A(z_{\max})$; from \cref{hellaby,eq:rho} we derive the prediction on $M(R_{\max},z_{\max})$ and $\rho(R_{\max},z_{\max})$. 

The absolute distance scale is set by two quantities. BAO measure distances in units of the sound horizon at the drag epoch, $r_d$. To express them in Mpc we use the \textit{Planck} 2018 value $r_d=147.09\pm0.26$~Mpc~\cite{Planck:2018vyg}, which is set by the physics before recombination and not by the late-time expansion we reconstruct. For each GP realization we draw $r_d$ from this distribution, so its uncertainty is propagated to the results. The value of $r_d$ does not affect $z_{\max}$, and it rescales $R_{\max}$ proportionally. For Pantheon+, the absolute scale is instead set by $h=0.7$. The effect of a different calibration is discussed below.

DESI DR2~\cite{DESIDR2II} measures $D_M/r_d$ and $D_H/r_d$, and in a spatially flat universe $D_H=\dd D_M/\dd z$. We therefore fit a single GP to $f=D_M/r_d$, with mean given by the $w_0w_a$ background above, and treat the $D_H/r_d$ data as measurements of its derivative $f'$, using the full DESI covariance. Since $D_A=f\,r_d/(1+z)$, the maximum lies where $(1+z)f'-f$ changes sign from positive to negative. The reconstruction is shown in Figure~\ref{fig:DA_DESI}. We find $z_{\max}=1.613\pm0.012$ and $R_{\max}=1773.4\pm6.1$~Mpc. This precision comes from the peak condition itself: at the maximum $R_{\max}\equiv D_A(z_{\max})=D_H(z_{\max})$, which is the quantity DESI measures directly. Moreover, since $\dd D_A/\dd z=0$ at the maximum, a shift $\delta z$ in $z_{\max}$ changes $R_{\max}$ only by $\mathcal{O}(\delta z^2)$.

We checked that our results do not depend on how the GP is used, on the choice of the mean, on the kernel, on the code, or on the calibration of the distances, as summerized below:

\paragraph*{\textbf{Trained and untrained GP.}}
In our main analysis the GP is \emph{trained} on the data, meaning that it is conditioned on the measurements, so that its realizations are drawn only among the functions allowed by the data and their covariance. In the \emph{untrained} approach of~\cite{Hwang:2022hla}, functions are instead drawn from the GP prior, independently of the data, and each function is then weighted by its $\chi^2$ against the data, computed with the full covariance. The data then only label the functions and never shape them. The two approaches give the same posterior~\cite{Hwang:2022hla}, but they compute it in completely different ways, so their agreement tests the whole reconstruction. Repeating the analysis with the untrained GP, we find $z_{\max}$ and $R_{\max}$ within $0.1\sigma$ of the trained values, with uncertainties within about 10\%. The untrained approach is much less efficient, and with a zero mean it does not converge, since functions drawn at random around zero almost never fit the data~\cite{Hwang:2022hla}.

\paragraph*{\textbf{Mean.}}
We tested different choices for the mean, to ultimately select the most conservative one. With a $\Lambda$CDM mean, with $\Omega_{\rm m}$ fixed to 0.3 or free, the shift in $R_{\max}$ is less than 1$\sigma$, while the uncertainty on $z_{\max}$ shrinks by about a factor of two. Fixing the shape of the mean artificially reduces the uncertainties~\cite{Hwang:2022hla}, which is why we present the results obtained by marginalizing over $(\Omega_{\rm m},w_0,w_a)$.

\paragraph*{\textbf{Kernel.}}
We use the squared-exponential kernel for simplicity. Matérn kernels give indistinguishable results, as also found in~\cite{Hwang:2022hla}.

\paragraph*{\textbf{Sampler.}}
The results in the main text are obtained with \texttt{emcee}~\cite{Foreman-Mackey:2012any}, with the mean computed analytically for a flat universe containing matter and $w_0w_a$ dark energy, neglecting radiation. We repeated them with \texttt{cobaya}~\cite{Torrado:2020dgo}, computing the mean with \texttt{class}~\cite{Blas:2011rf}, including radiation and one massive neutrino of $0.06$~eV. The two means differ by less than $4\times10^{-4}$, and the results are unchanged.

\paragraph*{\textbf{Calibration.}}
We repeated the BAO analysis with $r_d=136.9\pm2.1$~Mpc, in agreement with SH0ES results, i.e.\ with the local distance ladder rather than from the CMB. This value is approximately $6.9\%$ lower than the one used in the main text, so we expect $R_{\max}$ to decrease by the same amount, since $R_{\max}\propto r_d$. Similarly, for Pantheon+ we set $h=0.68$ instead of $0.7$, which should increase $R_{\max}\propto1/h$ by roughly $2.9\%$. We obtain, indeed $R_{\max}=1654.1\pm25.7$~Mpc for DESI ($-6.7\%$ from the main text result) and $R_{\max}=1763.3\pm39.4$~Mpc for Pantheon+ ($+2.9\%$ from the main text result). The change in $z_{\max}$ is within $0.6\sigma$ from the main text result.

\subsection{Full list of results and uncertainties}
\label{SM:2}
In Table~\ref{tab:DA_max_w0wa} we report the full set of 68$\%$~C.L. constraints obtained with the analyses discussed in the main text.

\begin{table*}[h]
\centering
\caption{Summary of 68\% C.L. constraints on the MS mass of the universe, derived from model-independent measurements of $R_{\max}$ obtained from state-of-the-art BAO and SNe Ia data. The first row shows mean value and experimental uncertainty on the QSO data point at $z=1.484$, as obtained by the DESI Collaboration. The second and third rows refer to the results of our GP reconstructions. DESI measurements are calibrated with $r_d=147.09\pm0.26$ Mpc, while the Pantheon+ catalog is anchored at $h=0.7$. The $\Lambda$CDM prediction for $\rho_{\rm{\scriptscriptstyle FLRW}}(z_{\max})$ corresponds to $\Omega_{\rm m} = 0.3$ and $h = 0.7$.}
\label{tab:DA_max_w0wa}
\begin{tabular}{lccccc}
\hline\hline
\rule{0pt}{3ex}
 & $R_{\max}$ & $z_{\max}$ & $M(R_{\max},z_{\max})$ & $\rho(R_{\max},z_{\max})$ & $\rho_{\rm{\scriptscriptstyle FLRW}}(z_{\max})$ \\[2pt]
 & [Mpc] & & [$10^{22}\,M_\odot$] & [$10^{-27}\,\mathrm{kg\,m^{-3}}$] & [$10^{-27}\,\mathrm{kg\,m^{-3}}$]\\[4pt]
\hline\rule{0pt}{3ex}
DESI QSO data point & $1807\pm45$ & $1.484$ & 
$1.888\pm0.047$ & $51.7\pm2.6$ & -- \\[4pt]
\hline\rule{0pt}{3ex}
DESI DR2 BAO (GP) & $1773.4\pm6.1$ & $1.613\pm0.012$ & $1.853\pm0.006$ & $53.68\pm0.37$ & -- \\[4pt]
~Pantheon+ (GP) & $1713\pm38$ & $1.58\pm0.11$ & $1.790\pm0.040$ & $57.5\pm2.6$ & -- \\[4pt]
\hline\rule{0pt}{3ex}
$\Lambda$CDM ($\Omega_{\rm m} = 0.3$, $h = 0.7$) & $1747.6$ & $1.605$ & $1.826$ & -- & $55.28$ \\[4pt]
\hline\hline
\end{tabular}
\end{table*}

\end{document}